\documentclass[aps,preprint,preprintnumbers,amsmath,amssymb,nofootinbib]{revtex4}
\usepackage{graphicx}
\usepackage{epsfig}
\begin{document}
\title{Testing Extended General Relativity with galactic sizes and compact gravitational sources.}
\author{$^{1,2}$ Mariano Anabitarte\footnote{E-mail address:
anabitar@mdp.edu.ar}, $^{1,2}$ Mat\'{\i}as Reynoso\footnote{E-mail
address: mreynoso@mdp.edu.ar}, $^{1}$ Pablo Alejandro
S\'anchez\footnote{E-mail address: pabsan@mdp.edu.ar}, $^{1,2}$
Jes\'us Mart\'{\i}n Romero\footnote{E-mail address:
jesusromero@conicet.gov.ar} and $^{1,2}$ Mauricio Bellini
\footnote{E-mail address: mbellini@mdp.edu.ar} }
\address{$^1$ Departamento de F\'isica, Facultad de Ciencias Exactas y
Naturales, Universidad Nacional de Mar del Plata, Funes 3350, C.P.
7600, Mar del Plata, Argentina.\\
$^2$ Instituto de Investigaciones F\'{\i}sicas de Mar del Plata (IFIMAR), \\
Consejo Nacional de Investigaciones Cient\'ificas y T\'ecnicas
(CONICET), Argentina.}

\begin{abstract}
Considering a five-dimensional (5D) Riemannian spacetime with a
particular stationary Ricci-flat metric, we have applied the
recently developed Extended General Relativistic theory to
calculate the lower limit for the size of some relevant galaxies:
the Milky Way, Andromeda and M87. Our results are in very good
agreement with observations.
\end{abstract}
\maketitle

\section{Introduction, basic equations and motivation}

The standard 4D General Relativity and its Newtonian weak-field
limit fail at describing the observed phenomenology when it is
applied to cosmic structure on galactic and larger scales. To
reconcile the theory with observations we need to assume that
$\sim\, 85\,\%$ of the mass is seen only through its observational
effect and that $\sim\, 74\,\%$ of the energy content of the
universe is due to either to an arbitrary cosmological constant or
to a not well defined dark energy fluid. The cosmological constant
problem appears to be so serious as the dark matter problem. The
Einstein equations admit the presence of an arbitrary constant
$\Lambda$.  Since observations indicate $\Lambda >0$, the dark
energy fluid has negative pressure. Current observations suggest
$\omega =-1$ at all probed epochs\cite{omega}, so models more
sophisticated than a simple $\Lambda$ could seem in principle
unnecessary.

In a previous paper\cite{mb} we developed a extended general
relativistic formalism from which we obtained an effective 4D
static and spherically symmetric metric which give us ordinary
gravitational solutions on small (planetary and astrophysical)
scales, but repulsive (anti gravitational) forces on very large
(cosmological) scales with $\omega =-1$. Our approach is an
unified manner to describe dark energy, dark matter and ordinary
matter in the framework of the induced matter theory\cite{im}. In
this letter we extend our calculations to study the lower limit
for the size of some galaxies (the Milky Way, Andromeda and
Messier 87) of interest.

\subsection{5D massive test particles dynamics}

In a previous work\cite{mb} we have considered a 5D extension of
General Relativity such that the effective 4D gravitational
dynamics has a vacuum dominated, $\omega =-1$, equation of state.
The starting 5D Ricci-flat metric $g_{ab}$, there considered is
determined by the line element\cite{mb,rom}
\begin{equation}\label{a1}
dS^{2}=\left(\frac{\psi}{\psi_0} \right)^{2}\left[ c^{2}f(r)dt^{2}
- \frac{dr^2}{f(r)}-r^{2}\left(d\theta ^{2}+sin^{2}(\theta)
\,d\phi ^{2}\right)\right]-d\psi^{2},
\end{equation}
where
$f(r)=1-(2G\zeta\psi_{0}/(rc^2))[1+c^{2}r^{3}/(2G\zeta\psi_{0}^{3})]$
is a dimensionless function, $\lbrace t,r,\theta,\phi\rbrace$ are
the usual local spacetime spherical coordinates employed in
general relativity and $\psi$ is the space-like extra dimension
that following the approach of the induced matter theory, will be
considered as non-compact. Furthermore, the space-like coordinates
$\psi$ and $r$ have length units, meanwhile $\theta$ and $\phi$
are angular coordinates, $t$ is a time-like coordinate and $c$
denotes the speed of light. We shall consider that $\psi_0$ is an
arbitrary constant with length units and the constant parameter
$\zeta$ has units of $(mass)(length)^{-1}$.

For a massive test particle outside of a spherically symmetric
compact object in 5D with exterior metric given by (\ref{a1}) the
5D Lagrangian can be written as
\begin{equation}\label{b1}
^{(5)}L=\frac{1}{2}g_{ab}U^{a}U^{b}=\frac{1}{2}\left(\frac{\psi}{\psi_0}\right)^{2}\left[c^{2}f(r)
\left(U^t\right)^2-\frac{\left(U^r\right)^2}{f(r)}-r^{2}\left(U^{\theta}\right)^2-r^{2}sin^{2}
\theta\left(U^{\phi}\right)^2\right] -
\frac{1}{2}\left(U^{\psi}\right)^2.
\end{equation}
We shall take $\theta=\pi/2$. Since $t$ and $\phi$ are cyclic
coordinates, hence their associated constants of motion $p_{t}$
and $p_{\phi}$, are
\begin{eqnarray}\label{b2}
p_{t}\equiv\frac{\partial\,^{(5)}L}{\partial U^t}&=&c^{2}\left(\frac{\psi}{\psi_0}\right)^{2}f(r)U^t,\\
\label{b3} p_{\phi}\equiv\frac{\partial\,^{(5)}L}{\partial
U^{\phi}}&=&-\left(\frac{\psi}{\psi_0}\right)^{2}r^{2} U^{\phi}.
\end{eqnarray}
Using the constants of motion given by (\ref{b2}) and (\ref{b3}),
we can express the five-velocity condition as follows:
\begin{equation}\label{b5}
\left(\frac{\psi_0}{\psi}\right)^{2}\frac{p_{t}^{2}}{c^{2}f(r)}-\left(\frac{\psi}{\psi_0}\right)^{2}\frac{\left(
U^{r}\right)^{2}}{f(r)}-\frac{p_{\phi}^2}{r^2}\left(\frac{\psi_0}{\psi}\right)^{2}-\left(
U^{\psi}\right)^{2}=c^{2}.
\end{equation}
After rearranging some terms and using the expression for $f(r)$,
the equation (\ref{b5}) can be written as
\begin{equation}\label{b6}
\frac{1}{2}\left(
U^{r}\right)^{2}+\frac{1}{2}\left(\frac{\psi_0}{\psi}\right)^{2}\left(
U^{\psi}\right)^{2} + V_{eff}(r) = E.
\end{equation}
If we identify the energy $E$ as
\begin{equation}\label{b8}
E=\frac{1}{2}\left(\frac{\psi_0}{\psi}\right)^{4}(p_{t}^{2}c^{-2}+p_{\phi}^{2}\psi_{0}^{-2})-
\frac{c^2}{2}\left(\frac{\psi_0}{\psi}\right)^{2}.
\end{equation}
the effective 5D potential $V_{eff}(r)$ results to be
\begin{eqnarray}
V_{eff}(r)&=&-\left(\frac{\psi_0}{\psi}\right)^{2}\frac{G\zeta\psi
_0}{r}+\left(\frac{\psi_0}{\psi}\right)^{4}\left[\frac{p_{\phi}^2}{2r^2}-\frac{G\zeta\psi
_0 p_{\phi}^2}{c^{2}r^3}\right] \nonumber \\
&-&
\frac{1}{2}\left(\frac{\psi_0}{\psi}\right)^{2}\left[\left(U^{\psi}\right)^2\left(\frac{2G\zeta\psi_0}{c^2
r}-\frac{r^2}{\psi_0^2}\right)-
\left(\frac{rc}{\psi_0}\right)^{2}\right]. \label{b7}
\end{eqnarray}
However, we are interested in the study of this potential for
massive test particles on static foliations $\psi=\psi_0=c/H_0$,
such that the dynamics evolves on an effective 4D manifold
$\Sigma_0$. From the point of view of an relativistic observer,
this implies that $U^{\psi}=0$.

\subsection{Physics on the 4D manifold $\Sigma_0$}

When we take a foliation $\lbrace\Sigma _{0} :\psi=\psi
_{0}\rbrace$ on (\ref{a1}), we obtain the induced metric given by
the 4D line element
\begin{equation}\label{a2}
dS^{2}_{ind}=c^{2}f(r)dt^{2}-\frac{dr^{2}}{f(r)}-r^{2}\,\left[d\theta
^{2}+sin^{2}(\theta) \,d\phi ^{2}\right],
\end{equation}
which is known as the Schwarzschild-de Sitter metric. From the
relativistic point of view, observers that are on $\Sigma_0$ move
with $U^{\psi}=0$. We assume that the induced matter on $\Sigma_0$
can be globally described by a 4D energy momentum tensor of a
perfect fluid $T_{\alpha\beta}=(\rho c^2
+P)U_{\alpha}U_{\beta}-Pg_{\alpha\beta}$, where $\rho(t,r)$ and
$P(t,r)$ are respectively the energy density and pressure of the
induced matter, such that
\begin{equation}\label{a7}
P= -\rho c^2 =-\frac{3c^4}{8\pi G}\frac{1}{\psi _{0}^2},
\end{equation}
which corresponds to a vacuum equation of state. We associate the
energy density of induced matter $\rho$ to a mass density of a
sphere of physical mass $m\equiv \zeta\psi_0$ and radius $r_0$. If
we do that, it follows that $m$ and the radius $r_0$ of such a
sphere are related by the expression
$\zeta=r^{3}_{0}/(2G\psi_{0}^3)$, such that $G\zeta = \sqrt{3}/ 9$
and there is only a single Schwarzschild radius. In this case the
Schwarzschild radius is $r_{Sch} = {\psi_0/\sqrt{3}} \ge r_0$.
When it is greater than the radius of the sphere of parameter
$\zeta$, the compact object has properties very close to those of
a black hole on distances $1 \gg r/\psi_0 > r_{Sch}/\psi_0$. This
condition holds when $G\zeta \le 1/ (2 \sqrt{27})\simeq 0.096225$.
For $G\zeta > \sqrt{3}/9$ one obtains $f(r)<0$ and there is not
Schwarzschild radius. When $G\zeta \le \sqrt{3}/ 9$ there are two
Schwarzschild radius, an interior $r_{S_i}$ and an exterior one
$r_{S_e}$, such that by definition $f(r_{S_i})=f(r_{S_e})=0$. In
this paper we shall focus in this last possibility which is
relevant for astrophysical scales. We shall assume that we live on
the 4D hypersurface $\Sigma _{H_0}:\psi _{0}=cH_{0}^{-1}$, $H_0$
being $H_{0}=73\,\frac{km}{sec}Mpc^{-1}$ the present day Hubble
constant.

When one takes $U^{\psi}=0$, the induced potential $V_{ind}(r)$ on
the hypersurface $\Sigma _0$ is given by
\begin{equation}\label{b9}
V_{ind}(r)=-\frac{G m}{r}+\frac{p_{\phi}^{2}}{2r^2}-\frac{G
m}{c^2}\frac{
p_{\phi}^2}{r^3}-\frac{c^2}{2}\left(\frac{r}{\psi_0}\right)^{2},
\end{equation}
where $m=\zeta\psi _0$ is the effective 4D physical mass. The
confining force $\Phi^{\psi}=\epsilon/\psi_0$ is perpendicular to
the penta-velocities $U^{\mu}$ on all the hypersurface $\Sigma_0$,
so that the system is conservative on $\Sigma_0$. Hence
$\Phi^{\psi}$ cannot be interpreted as a fifth force.

The first two terms in the right hand side of (\ref{b9})
correspond to the classical potential, the third term is the usual
relativistic contribution and the last term is a new contribution
coming from the 5D metric solution (\ref{a1}). The acceleration
associated to the induced potential (\ref{b9}) reads
\begin{equation}\label{in2}
a=-\frac{Gm}{r^2}+\frac{p_{\phi}^2}{r^3}-\frac{3Gm}{c^2}\frac{p_{\phi}^2}{r^4}+\frac{rc^{2}}{\psi
_0^2}.
\end{equation}

The condition for circular motion of the test particle
$(dV_{ind}/dr)=0$ acquires the form
\begin{equation}\label{b10}
r^{5}-\frac{Gm}{c^2}\psi_{0}^{2}r^{2}+\frac{p_{\phi}^{2}\psi_{0}^2}{c^2}r-\frac{3Gm}{c^4}p_{\phi}^{2}\psi_{0}^{2}=0.
\end{equation}
By expressing the equation (\ref{b6}) as a function of the angular
coordinate $\phi$ (indeed assuming $1/u=r=r(\phi)$), we obtain,
after make the derivative with respect to $\phi$, the orbit
equation on 4D
\begin{equation}\label{b13}
\frac{d^{2}u}{d\phi^2}+u+c^{2}p_{\phi}^{-2}\psi_{0}^{-2}u^{-3}=\left(\frac{G
m}{c^2}\right) p_{\phi}^{-2}+\left(\frac{3G m}{c^2}\right) u^{2}.
\end{equation}
This equation is almost the same that the one usually obtained in
the 4D general theory of relativity for an exterior Schwarzschild
metric with the exception of the third term on the left hand side.
This new term could be interpreted as a new contribution coming in
this case from the extra coordinate.

\subsection{Motivation}

It is well known that galaxies are finite in size. However, from
the gravitational point of view this fact cannot be explained in a
satisfactory manner. From the observational point of view it is
considered that the limit of one galaxy is where the angular velocity of
matter around the center is zero: $\dot\phi=0$. At this radius,
the squared momentum $p^2_{\phi}$
\begin{equation}\label{p_}
p^2_{\phi} =  \frac{c^2}{(c/H_0)^2} \frac{\left(G m (c/H_0)^2 -
r^3 c^2\right)}{\left(c^2 r - 3 G m\right)},
\end{equation}
becomes zero. Hence, we shall consider some examples of galaxies
to determine their size once we know the masses of the BH which
are in the center of each one of these galaxies. This radius will
be that for which $p_{\phi}(r_{size})=0$ in eq. (\ref{p_}):
\begin{equation}
r_{size} = \left(\frac{G m}{H^2_0}\right)^{1/3}.
\end{equation}
However, this radius should be only a lower limit for the galactic
size, because in our calculation we are neglecting the galactic
mass outside the BH.

In in this letter, we shall deal with size of galaxies in the
present day, so that we shall consider that $(c/H_0)$ is the
present day Hubble horizon. However, for very distant galaxies one
should take into account the horizon of the galaxy at the moment
of the signal emission.

\section{Some examples}

In this letter we are interested to study galaxies to estimate its
size once we know the compact object [or black hole (BH)] which is
in its center. This kind of galaxies has spiral arms which evolves
on the plane perpendicular to the angular moment.

\subsection{Milky Way}

\begin{figure*}
\includegraphics[height=15cm]{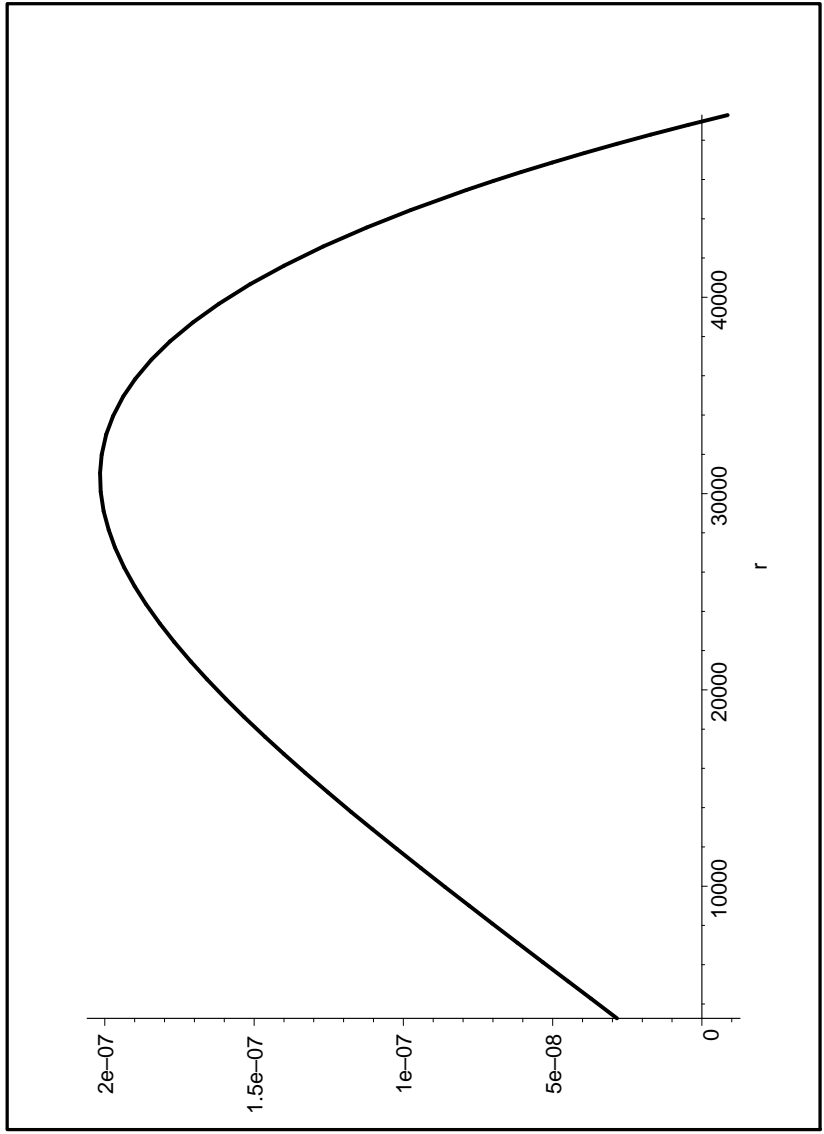}\caption{\label{figura1} $p^2_{\phi}$ as a function of the Milky Way galactic radius $r_{MW}$ [in
light years ($ly$)] for a BH mass $4.1 \times
10^{6}\,M_{\bigodot}$.}
\end{figure*}

The immediate case to be studied is our galaxy (or Milky Way),
which has in its center a BH of a mass $\sim 4.1 \times 10^6 \,\,
M_{\bigodot}$\cite{reid}. It is agreed that the Milky Way is a
barred spiral galaxy, with observations suggesting that it is a
spiral galaxy. The stellar disk of the Milky Way galaxy is
approximately $10^5 \,ly$ in diameter, and is considered to be, on
average, about $10^3 \, ly$ thick. It is estimated to contain at
least 200 billion stars and possibly up to $400$ billion stars,
the exact figure depending on the number of very low-mass, or
dwarf stars, which are hard to detect, especially more than $300\,
ly$ from the Sun, and so current estimates of the total number
remain highly uncertain, though often speculated to be around
$250$ billion. In the figure (\ref{figura1}) we plot the evolution
of $p^2_{\phi}$ as a function of the galactic radius. It is
obvious from the graphic that the lower limit of the radius of our
galaxy is close to $5 \times 10^4 \, ly$, in agreement with
observations.

The Galactic Halo extends outward, but is limited in size by the
orbits of two Milky Way satellites, the Large and the Small
Magellanic Clouds, whose distance is at about $1.8 \times
10^5\,\,ly$ \cite{connors}. In our model it corresponds with a
galactic mass of $2 \times 10^8\,\, M_{\bigodot}$, which is very
close to the Milky Way mass.

\subsection{Andromeda}

\begin{figure*}
\includegraphics[height=15cm]{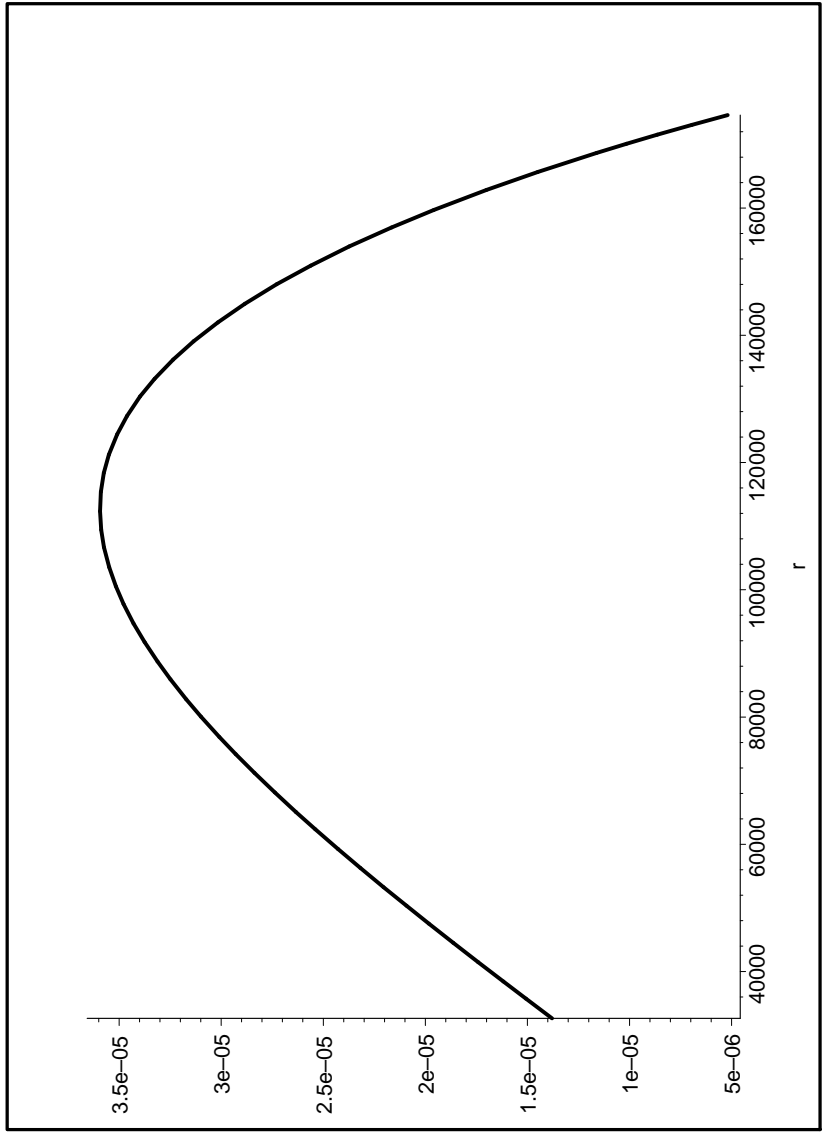}\caption{\label{figura2} $p^2_{\phi}$ as a function of the Andromeda galactic radius $r_A$ [in
light years ($ly$)] for three different BH masses: $3.0 \times
10^{7}\,M_{\bigodot}$ with thick continuous line, $6.2 \times
10^{7}\,M_{\bigodot}$ with thin continuous line and $1.0 \times
10^{8}\,M_{\bigodot}$ with points.}
\end{figure*}

Another case of interest is the Andromeda galaxy. It is also known
as Messier 31, {\bf M31} (or {\bf NGC 224}). Andromeda is the
nearest spiral galaxy to the Milky Way. {\bf M31} was the second
galaxy in which stellar dynamics revealed the presence of a
supermassive black hole\cite{Kormendy}. Axisymmetric dynamical
models implied BH masses of $(1-10) \times 10^7\, M_{\bigodot}$.
The smallest masses were given by disk models, and the largest
were given by spherical models. Hubble Space Telescope (HST)
spectroscopy reveals a dark mass (presumed black hole) located at
the center of this cluster with an estimated mass of $ 1.4 \pm
^{0.7}_{0.3} \times 10^8\,\, M_{\bigodot}$\cite{Bender}. In the
figure (\ref{figura2}) we show the evolution of $p^2_{\phi}$ for
three different BH mass values ($3.0 \times 10^{7}\,M_{\bigodot}$
with thick line, $6.2 \times 10^{7}\,M_{\bigodot}$ with thin line
and $1.0 \times 10^{8}\,M_{\bigodot}$ with points), as a function
of the minimum galactic radius. From the graphic we infer that the
radius of Andromeda is greater than $1.2 \times 10^5 \, ly$ for a
BH mass greater than $6.2 \times 10^{7}\,M_{\bigodot}$. Since the
observations agree with a minimum galactic radius close to $1.2
\times 10^5 \, ly$, we infer that the BH mass in the core of
Andromeda would have a mass close to $6.2 \times
10^{7}\,M_{\bigodot}$.

\subsection{Messier 87}

\begin{figure*}
\includegraphics[height=15cm]{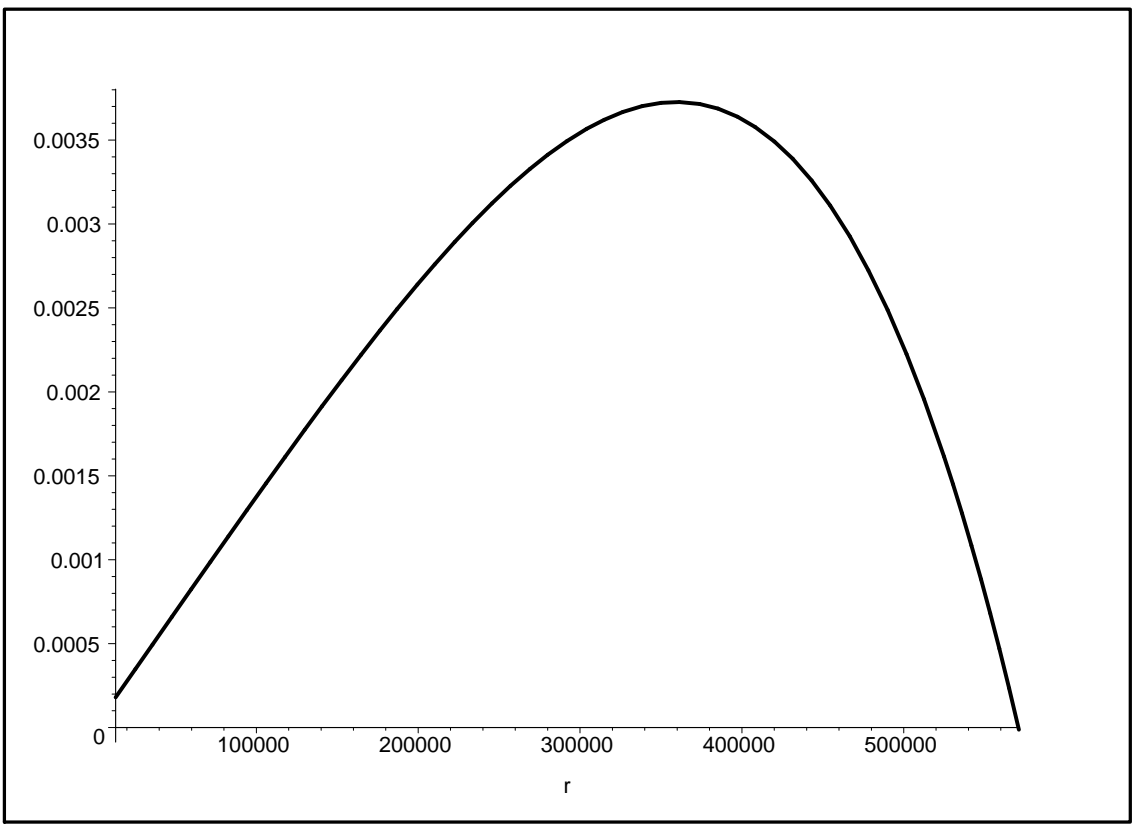}\caption{\label{figura3} $p^2_{\phi}$ as a function of the M87 galactic radius $r_{M87}$ [in light years
($ly$)] for two different BH masses: $2.4 \times
10^{9}\,M_{\bigodot}$ with points and $6.0 \times
10^{9}\,M_{\bigodot}$ with continuous line.}
\end{figure*}

A very interesting example is the galaxy Messier 87 (or {\bf
M87}). At the core is a supermassive BH with mass estimated in the
range $(2.4 - 6.5) \times 10^{9}\,\,M_{\bigodot}$\cite{Takahashi}.
{\bf M87} forms the primary component of an active galactic
nucleus that is a strong source of multiwavelength radiation,
particularly radio waves. A jet of energetic plasma originates at
the core and extends out at least 5000 ly\cite{Baade}. In the
figure (\ref{figura3}) it is shown that its bound radius $r_{M87}$
is really bigger than other galaxies, $r_{M87}$ being in the range
$(4.05 \times 10^5  < r_{M87} < 5.65 \times 10^5) \, ly $.

\section{Final Comments}

We have calculated the lower radius for three different galaxies
(the Milky Way, Andromeda and Messier 87) using the estimation for
the BH masses in their core of these ones. Our results agree very
good with observations. In view of our calculations we conclude
that the mass of the BH in the Milky Way should be close to $\sim
4.1 \times 10^6 \,\, M_{\bigodot}$  for a minimum radius close to
$r_{MW} \simeq 4.9 \times 10^{4} \, ly$. Moreover, taking into
account that our galaxy is limited in size by the orbits of two
Milky Way satellites (whose distance is at about $1.8 \times
10^5$), we conclude that our galaxy mass should take a mass of at
least $2 \times 10^8\,\, M_{\bigodot}$.

In the case of Andromeda, we conclude that the mass of the BH in
its center should be close to $r_{A} \simeq 6.2 \times 10^{7} \,\,
M_{\bigodot}$ in order to obtain a minimum galactic size close to
$r_A \simeq 1.2 \times 10^{5} \, ly$. Finally, for the range of
masses $(2.4 - 6.5) \times 10^{9}\,\,M_{\bigodot}$, our
calculations conclude that the bound radius $r_{M87}$, of Messier
87 should be in the range $(4.05 \times 10^5 < r_{M87} < 5.65
\times 10^5) \, ly $.

Overall, we have shown how the here adopted theory of extended
General Relativity implies the existence of the radius $r_{size}$
for a given gravitational source. Beyond this radius the
gravitational force becomes repulsive. As discussed above for the
three example cases, this radius is yet compatible with the
observed size of the source.

\section*{Acknowledgements}

\noindent The authors acknowledge UNMdP and CONICET Argentina for
financial support.

\end{document}